\documentclass[aps,prl,floatfix,twocolumn,footinbib]{revtex4}
\usepackage{amsmath}
\usepackage{epsfig}
\usepackage{amsfonts}
\usepackage{amssymb}
\usepackage{graphicx}
\begin{document}
\title{A Proposal to Measure the Quasiparticle Poisoning Time of Majorana Bound States}
\author{Jacob R. Colbert and Patrick A. Lee}
\affiliation{Department of Physics, Massachusetts Institute of Technology, Cambridge, MA 02139, USA}
\begin{abstract}
We propose a method of measuring the fermion parity lifetime of Majorana fermion modes due to quasiparticle poisoning. We model quasiparticle poisoning by coupling the Majorana modes to electron reservoirs, explicitly breaking parity conservation in the system. This poisoning broadens and shortens the resonance peak associated with Majorana modes. In a two lead geometry, the poisoning decreases the correlation in current noise between the two leads from the maximal value characteristic of crossed Andreev reflection. The latter measurement allows for calculation of the poisoning rate \emph{even if temperature is much higher than the resonance width}. 
\end{abstract}
\maketitle
The promise of topologically robust quantum computation has been a major motivation in condensed matter physics over the past decade. In such schemes quantum information is not stored locally but is stored in a global state of the system. In this way systems are protected against decoherence by local perturbations. A simple and potentially realizable platform for non-local quantum information storage is in systems with Majorana fermions, which split a single fermionic mode into two spatially separated Majorana bound states.

Majorana bound states are defined by the operator algebra: $\gamma_i^\dagger=\gamma_i$ and $\{\gamma_i,\gamma_j\}=2\delta_{ij}$ \cite{Kitaev-2001}. They have been theorized to exist in many different condensed matter systems, including 1D superconductors with p-wave pairing\cite{Kitaev-2001}, 2D $p_x+ip_y$ superconductors\cite{Read-2000}, topological insulator/superconductor heterostructures\cite{Fu-2008}, and semiconductor/superconductor heterostructures\cite{Lutchyn-2010,Oreg-2010}. There has been some recent success in the lattermost proposal. Recent experimental results in semiconductor nanowires show a zero-bias conductance peak, a potential indicator of Majorana modes\cite{Mourik-2012,Deng-2012,Das-2012,Churchill-2013} intensifying the interest in the field. The recent experiments, however, have not yet reached a low enough temperature to see the theoretically predicted quantized peak\cite{Bolech-2007,Law-2009} and other non-topological explanations of the peak have been suggested \cite{Liu-2012}. 

Systems with Majorana qubits are only protected under perturbations that preserve fermion parity; that is, they only involve the transfer of Cooper pairs \cite{Nayak-2008}. Perturbations that switch the fermion parity of the system, involving unpaired electrons, dubbed quasiparticle poisoning, will change the state of a Majorana qubit. The time scale of this poisoning rate is then a limiting factor for performing quantum computations. Recent theoretical calculations show that this poisoning rate may be problematically large for performing adiabatic gate operations\cite{Rainis-2012}. 

In light of this challenge, it is essential to be able to measure this poisoning rate. There have been several proposals to measure the rate based on SQUIDs in topological superconductor/superconductor heterosturctures\cite{Lutchyn-2010,Fu-2009}, a quantum dot coupled to a topological superconducting wire\cite{Leijnse-2011}, or direct measurement of parity relaxation times\citep{Burnell-2013}. In this article we propose a relatively simple experimental setup that doesn't require an interference measurement, based on the two lead transport experiments proposed by Nilsson et al.\cite{Nilsson-2008} and Liu et al.\cite{Liu-2013}. Our proposed measurement gives a direct probe of the breakdown of non-locality due to quasiparticle poisoning.

\begin{figure}
\includegraphics[width=\columnwidth]{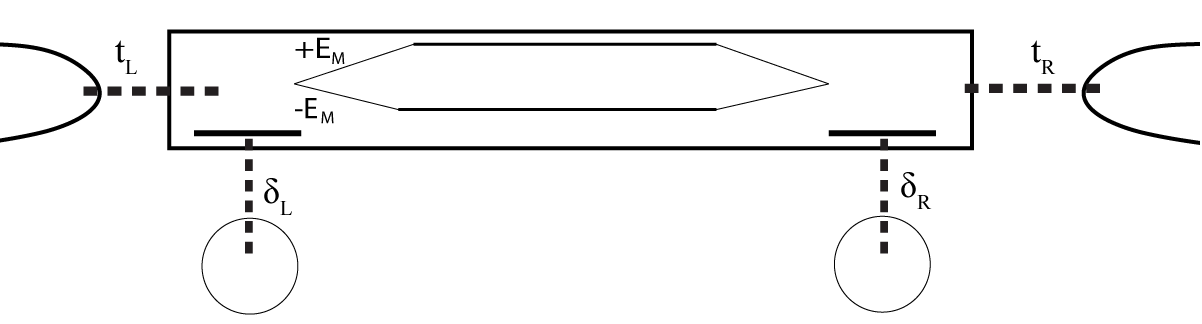}
\caption{Sketch of the experimental setup: Two tunneling current leads are connected to the ends of a grounded superconducting wire which support Majorana bound states at its ends. The Majoranas are split by energy 2$E_M$. In addition reservoirs are coupled to the Majorana bound states with matrix elements $\delta_L$ and $\delta_R$ to simulate the effect of quasiparticle tunneling.}
\label{fig:setup}
\end{figure}
We consider the same experimental geometry as in \cite{Nilsson-2008} and \cite{Liu-2013}: we have a grounded topological superconducting wire, with each Majorana bound state coupled to a normal lead. We model quasiparticle poisoning by coupling fermion reservoirs to each Majorana mode. We consider the limit with $k_B T,eV\ll \Delta$, so that the Majorana modes are the only modes accessible to electrons tunneling from the leads. Explicitly, our effective Hamiltonian is 

\begin{equation}
H=H_0+iE_M\gamma_L\gamma_R+\sum_{\alpha=L,R}\big(t_\alpha\gamma_\alpha(c_\alpha+c_\alpha^\dagger)+\delta_\alpha\gamma_\alpha(f_\alpha+f_\alpha^\dagger)\big)
\label{eq:Heff}
\end{equation}
Here $$H_0=\sum_{\epsilon,\alpha=L,R}\big(\epsilon c_\alpha^\dagger(\epsilon) c_\alpha(\epsilon)+\epsilon f_\alpha^\dagger(\epsilon) f_\alpha(\epsilon)\big)$$ gives the lead and bath Hamiltonian, $E_M$ gives the splitting of the coupled Majorana modes, $t_\alpha$ gives the coupling to the leads with electron creation operators $c_\alpha^\dagger$ at the interface, and $\delta_\alpha$ gives the coupling to the quasiparticle baths with creation operators $f_\alpha^\dagger$ at the interface. We take the wide band limit for all leads and baths, and the results are independent of the details of the dispersions. $\gamma_\alpha=\gamma_\alpha^\dagger$ are the Majorana operators. Generally we can consider all couplings to be real by redefining the electron wave functions. We introduce the self-energies due to tunneling, $\Gamma_t^\alpha=2\pi N_\alpha(0) t_\alpha^2$, and  poisoning, $\Gamma_p^\alpha=2\pi N_\alpha(0) \delta_\alpha^2$, where $N_\alpha(0)$ is the position and energy independent density of states of the relevant reservoir. Throughout the paper we consider the case where the coupling between the MBSs is sufficiently strong that $E_M$ is much stronger $\Gamma$.

 The scattering matrix can be written using a generalization of the relation given by  Fisher and Lee: 
$$S_{\alpha\beta,ij,ab}(E)=1+i\sqrt{\Gamma_a^{\alpha}\Gamma^\beta_b}(H_M-E-Wg^rW^\dagger )^{-1}_{\alpha\beta}$$
 where $\alpha,\beta\in\{L,R\}$, $a,b\in\{t,p\}$ denote a lead or quasiparticle reservoir respectively, and $i,j\in\{e,h\}$ denote electron and hole channels.
$$W=\left(\begin{array}{cccccccc}
t_L & 0 & \delta_L & 0 & t_L & 0 & \delta_L & 0\\
0 & t_R & 0 & \delta_R & 0 & t_R & 0 & \delta_R \end{array}\right)$$
gives the coupling of the Majoranas, $\{\gamma_L,\gamma_R\}$, to the leads and to the reservoirs in the basis $\{c_L,c_R,f_L,f_R,c_L^\dagger,c_R^\dagger,f_L^\dagger,f_R^\dagger\}$,
$g^r$ is the surface Green's function for the leads and baths given by an $8$x$8$ diagonal matrix with entries $-i\pi N(0)$ for the relevant density of states,
 and $$H_M=\left(\begin{array}{cc}
0 & iE_M \\
-iE_M & 0 \end{array}\right)$$ gives the coupling between the pair of Majorana bound states. 

Since the only dependence on the electron vs. hole channel in the scattering matrix is in the identity matrix term, the scattering matrix can be written in the form $$S=\left(\begin{array}{cc}
1+A &  A \\
A  & 1+A \end{array}\right)$$

We can then, following \cite{Nilsson-2008}, write the current and current-current correlators in the form:
\begin{equation}
\bar{I_i}=\frac{2e}{h}\int^{eV}_0 (AA^\dagger)_{ii}dE
\label{eq:I}
\end{equation}
\begin{equation}
\begin{aligned}
C_{ij}&=\int^{+\infty}_{-\infty}\langle\delta I_i(0)\delta I_j(t)\rangle \\
&=e\bar{I_i}\delta_{ij}+\frac{2e^2}{h}\int^{eV}_0[|A_{ij}+(AA^\dagger)_{ij}|^2-|(AA^\dagger)_{ij}|^2]dE
\end{aligned}
\label{eq:C}
\end{equation}
where $\delta I_i(t)=I_i(t)-\bar{I_i}$. We also define $G_i=\frac{dI_i}{dV}$, the differential conductance, and $P_{ij}=\frac{dC_{ij}}{d(eV)}$, the differential contribution to the noise by electrons with energy $eV$.

We first consider the one lead case by setting $t_R$ to zero. With no poisoning, the differential conductance has a quantized $2e^2/h$ resonance peak characteristic of Majorana induced resonant Andreev reflection.\cite{Law-2009,Bolech-2007} Using the above relations we can expand the differential conductance near the maximum, $|E-E_M|\ll E_M$, obtaining 
\begin{equation}
G_L\approx\frac{2e^2}{h}\frac{\Gamma_t^L(\Gamma_t^L+\Gamma_p^L+\Gamma_p^R)}{4(E-E_M)^2+(\Gamma_t^L+\Gamma_p^L+\Gamma_p^R)^2}
\label{eq:ResonanceConductance}
\end{equation}
From this we can conclude that with poisoning the conductance peak is shortened from $2e^2/h$ by a factor of $\Gamma_t^L/(\Gamma_t^L+\Gamma_p^L+\Gamma_p^R)$. Its width is broadened from $\Gamma_t^L$ to $\Gamma_t^L+\Gamma_p^L+\Gamma_t^L$; the coupled Majorana modes can decay not only into the coupled lead, but also into the quasiparticle reservoirs. This result is given in Ref. \citenum{Liu-2013} and implicitly contained in Ref. \citenum{Bolech-2007}.  In view of the large conductance background in the current experimental data in Refs. \citenum{Mourik-2012,Deng-2012,Das-2012,Churchill-2013}, the poisoning time may be playing a strong role in limiting the height of the zero bias peak.

\begin{figure}
\includegraphics[width=\columnwidth]{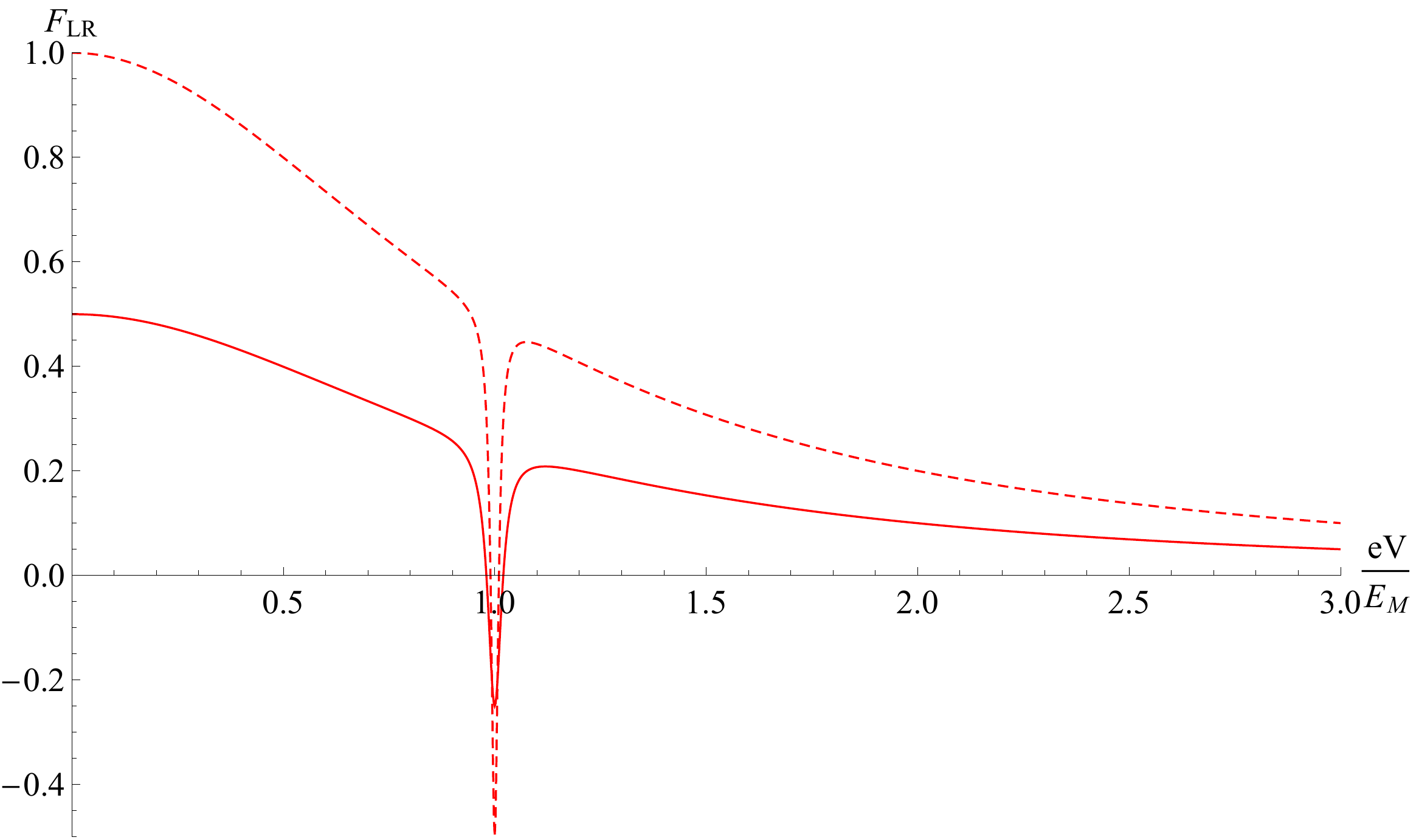}
\caption{The zero temperature cross-correlation differential Fano factor, $F_{LR}$ for $\Gamma_t^L=\Gamma_t^R=\Gamma$, and $E_M=100\Gamma$ without poisoning (dashed), and with poisoning rate $\Gamma_p^L=\Gamma_p^R=\Gamma$ (solid).}
\label{fig:FLR}
\end{figure}

\begin{figure}
\includegraphics[width=\columnwidth]{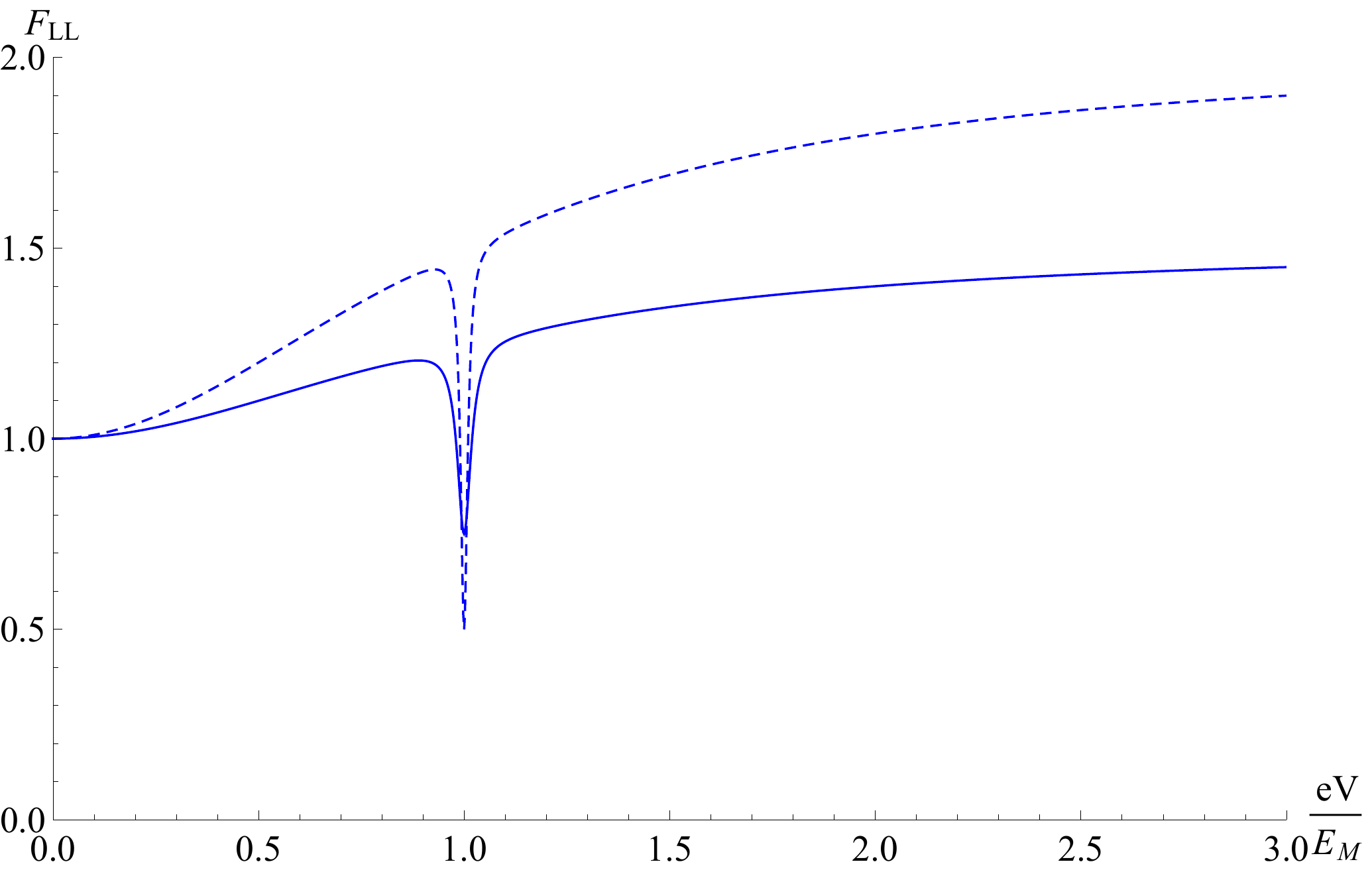}
\caption{The zero temperature differential Fano factors for the left lead, $F_{LL}$, $\Gamma_t^L=\Gamma_t^R=\Gamma$, and $E_M=100\Gamma$ without poisoning (dashed), and with poisoning rate $\Gamma_p^L=\Gamma_p^R=\Gamma$ (solid). }
\label{fig:FLL}
\end{figure}
The resonance peak measurement allow for the measurement of the poisoning rate $\tau_p=1/(\Gamma_p^{L}+\Gamma_p^{R})$. By measuring the height of the differential conductance peak we can measure the ratio $(\Gamma_p^L+\Gamma_p^R)/\Gamma_t$. The width of the peak is given by $\Gamma_t+\Gamma_p^L+\Gamma_p^R$. Given these two pieces of information we can calculate the poisoning rate. These results however are only valid in the zero temperature limit. At temperatures comparable to the tunneling scales, $\Gamma_t^\alpha$, thermal broadening becomes important and the height of the peak is reduced. In order to find a way of measuring the poisoning time that is robust for temperatures $k_BT>\Gamma_t$ we can look at properties away from the resonance: the  conductance and noise in the two terminal geometry in the regime $k_BT\ll eV\ll E_M$, and $k_BT> \Gamma_t$.

To simplify notations we begin by discussing the case when the tunneling rates to each side are equal, i.e. $\Gamma_t^L=\Gamma_t^R=\Gamma_t$ and $\Gamma_p^L=\Gamma_p^R=\Gamma_p$, and return to the general case later. We consider correlation in the current noise spectrum first in the zero temperature limit. This provides a more specific signature of the presence of Majoranas. With no quasiparticle poisoning and in the low voltage regime, the current is dominated by crossed Andreev reflection, where an incoming electron in the left lead is transmitted as an outgoing hole in the right lead.\cite{Nilsson-2008}. This can be seen by looking at the differential Fano factors, the ratio of the differential noise correlator to the differential conductance\cite{Law-2009}. For low temperatures, $k_BT\ll eV$, the noise is dominated by shot noise and the Fano factor can be interpreted as the charge transferred in each tunneling event. The Fano factor for tunneling from an individual lead, $F_{\alpha\alpha}=P_{\alpha\alpha}/G_\alpha$, is $1$ while the Fano factor for the total current, $F_{tot}=\sum_{ij}P_{ij}/\sum_i G_i$, is $2$ showing that in each tunneling event $2e$ charge enters the wire with $e$ coming from each side. Additionally the cross-correlation Fano factor, $F_{LR}=2P_{LR}/(G_L+G_R)$, is $1$ saturating the inequality $2|P_{LR}|\leq P_{LL}+P_{RR}$ for stochastic processes\cite{Nilsson-2008}.

Turning on poisoning quickly kills this correlation. Using equations \ref{eq:I} and \ref{eq:C} we can explicitly calculate the Fano factors in Mathematica.  A plot of the Fano factors comparing the cases with and without poisoning is shown in Figs. \ref{fig:FLR} and \ref{fig:FLL}. In the low voltage regime, we can expand the result in $\Gamma^2/E_M^2$ and $E^2/E_M^2$ and keep only the zeroth order term. The Fano factor for noise correlations falls as 
\begin{equation}
F_{LR}=\frac{2P_{LR}}{G_L+G_R}(eV=0)\approx\frac{\Gamma_t}{\Gamma_t+\Gamma_p}
\label{eq:FLR}
\end{equation}
This result is easily interpreted. Electrons are still added to the wire in pairs, but with poisoning the chance that the other electron came from the other lead rather than from a quasiparticle is $\frac{\Gamma_t}{\Gamma_t+\Gamma_p}$. In contrast the Fano factor, $F_{LL}$ for the current from an individual lead remains $e$, as it is still a single electron that tunnels through the barrier from the lead. The presence of poisoning simply opens up a new channel for where the other electron comes from; it could be from the other lead or from a quasiparticle reservoir. From Fig. \ref{fig:FLL} we see that for $eV>E_M$, $F_{LL}$ is reduced from $2$ to $1.5$ by quasi particle poisoning. However this feature does not depend on the presence of Majorana bound states and will happen as long as a quasiparticle decay channel is available in addition to Andreev reflection. Therefore we conclude that the single lead noise measurements of $F_{LL}$ and $F_{RR}$ are not sufficient to provide information on the poisoning times of Majorana bound states and we will focus on $F_{LR}$ below.

Away from perfect tuning of the tunneling amplitudes of the two leads, $\Gamma_t$ in Eq.\ref{eq:FLR} is replaced by 
\begin{equation}
\Gamma_t^{tot}=(\tau_{avg})^{-1}=(\frac{1}{2}(\tau_L+\tau_R))^{-1}=\frac{2\Gamma_t^L \Gamma_t^R}{\Gamma_t^L+\Gamma_t^R}
\end{equation} With unequal poisoning on each side $\Gamma_p$ is replaced by 
\begin{equation}\Gamma_p^{avg}=\frac{\Gamma_t^R}{\Gamma_t^L+\Gamma_t^R}\Gamma_p^L+\frac{\Gamma_t^L}{\Gamma_t^L+\Gamma_t^R}\Gamma_p^R
\end{equation}



The measurements for the Fano factors give the ratios of the poisoning time scales to the tunneling time scales. In order to calculate the value of the poisoning time we need another measurement that survives in the finite temperature case. This is given by the differential current at low bias. A generalization of the result given in Ref. \citenum{Liu-2013}, shows that at zero bias the differential conductance is given by 
\begin{equation}
G_L\approx\frac{2e^2}{h}\frac{\Gamma_t^L(\Gamma_t^{R}+\Gamma_p^{R})}{E_M^2}
\label{eq:LowBiasConductance}
\end{equation} and similarly for $G_R$. In addition to crossed Andreev reflection, electrons can enter in pairs one coming from the lead the other from the quasiparticle reservoir. By combining several measurements we can determine the poisoning rate (see Appendix B for details).

These measurements remain valid for temperatures $k_BT>\Gamma$ as they are measured away from the resonance. Away from resonance the differential noise and conductance change on an energy scale set by $E_M$. As long as $k_BT\ll E_M$, the thermal sampling of different points on the differential noise and conductance curves has little effect. To calculate the Fano factors and differential conductance at finite temperature we use the results given in Ref. \citenum{Anantram-1996}, given also in the appendix. In Fig. \ref{fig:temperature}, we show the temperature dependence of the Fano Factor at low voltage. We note that as long as there is a range of voltages where $k_BT\ll eV\ll \Delta$, Eqns. \ref{eq:FLR} and \ref{eq:LowBiasConductance} still hold.

\begin{figure}
\includegraphics[width=\columnwidth]{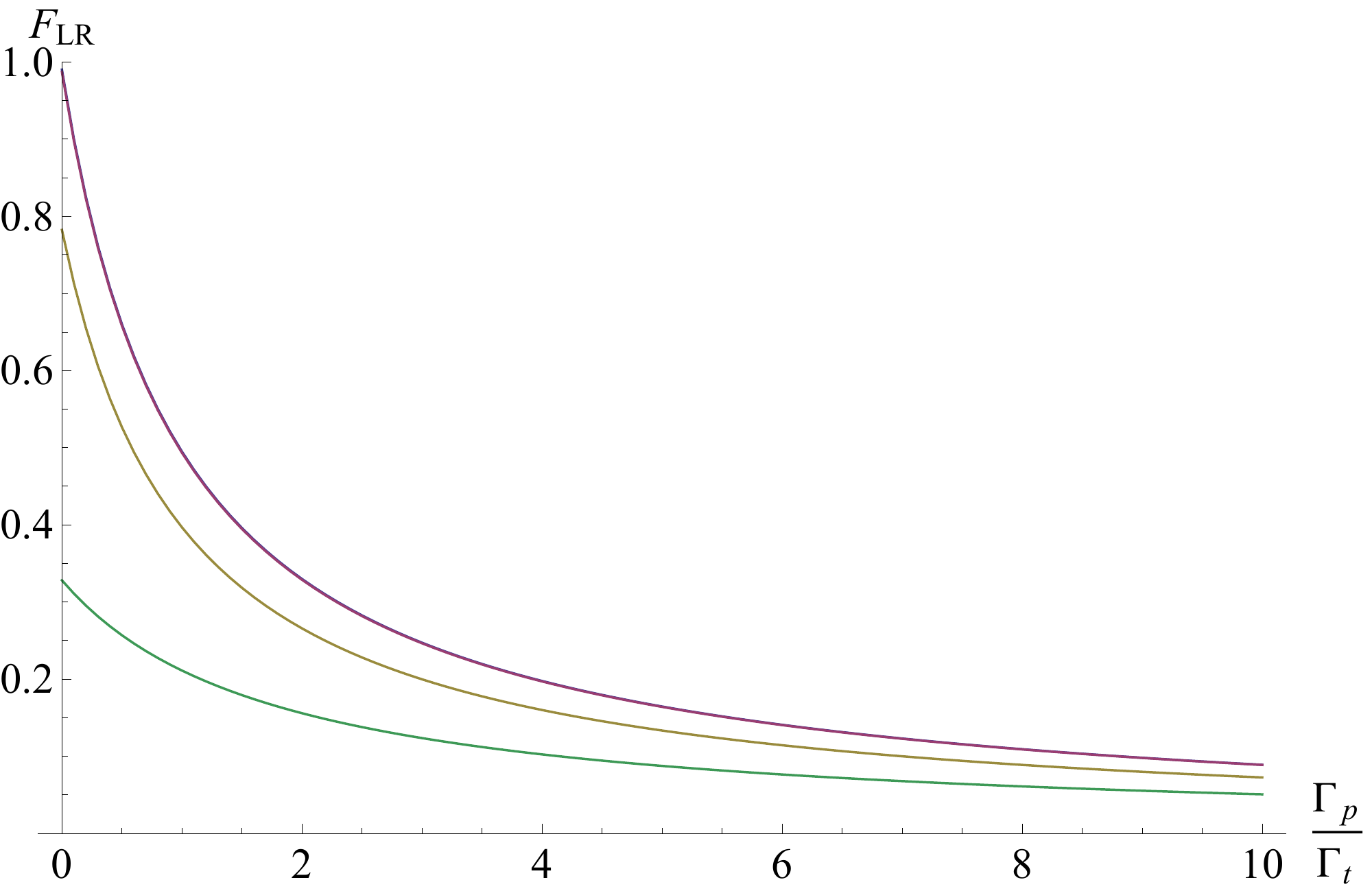}
\caption{The correlation differential Fano factor at fixed low bias $eV=10\Gamma_t<E_M=100\Gamma_t$ as a function of poisoning. We show curves at temperatures $T=0,4,8,12\Gamma$. Note that the curves for $T=0$ and $T=4\Gamma_t$ are nearly overlapping. Only when $k_BT\sim eV$ is the temperature effect strong.}
\label{fig:temperature}
\end{figure}

Throughout the paper we have been using the differential Fano factors, as they more clearly demonstrate the physical behavior of electrons at a given energy\cite{Liu-2013}, but we could instead use the more easily measurable Fano factor for the integrated noise and current. In the low voltage regime, the correction terms due to finite bias in both the differential conductance and noise go as $(eV)^2/E_M^2$. Ignoring these terms the current and noise are linear and the results given here hold for the Fano factors $\tilde{F}_{\alpha\alpha}=C_{\alpha\alpha}/(eI_\alpha)$ and $\tilde{F}_{LR}=2C_{LR}/e(I_L+I_R)$. In particular $\tilde{F}_{LR}$ is the same as shown in Fig. \ref{fig:temperature} and will be an equally effective measure.

PAL thanks the support of DOE grant DE-FG02-03-ER 46076 and the John Templeton Foundation. JC acknowledges the support of NSF Grant
No. DGE-0801525, \emph{IGERT: Interdisciplinary
Quantum Information Science and Engineering}.
\bibliographystyle{apsrev}
\bibliography{references}

\begin{thebibliography}{20}
\expandafter\ifx\csname natexlab\endcsname\relax\def\natexlab#1{#1}\fi
\expandafter\ifx\csname bibnamefont\endcsname\relax
  \def\bibnamefont#1{#1}\fi
\expandafter\ifx\csname bibfnamefont\endcsname\relax
  \def\bibfnamefont#1{#1}\fi
\expandafter\ifx\csname citenamefont\endcsname\relax
  \def\citenamefont#1{#1}\fi
\expandafter\ifx\csname url\endcsname\relax
  \def\url#1{\texttt{#1}}\fi
\expandafter\ifx\csname urlprefix\endcsname\relax\def\urlprefix{URL }\fi
\providecommand{\bibinfo}[2]{#2}
\providecommand{\eprint}[2][]{\url{#2}}

\bibitem[{\citenamefont{Kitaev}(2001)}]{Kitaev-2001}
\bibinfo{author}{\bibfnamefont{A.~Y.} \bibnamefont{Kitaev}},
  \bibinfo{journal}{Phys. Usp.} \textbf{\bibinfo{volume}{44}},
  \bibinfo{pages}{131} (\bibinfo{year}{2001}).

\bibitem[{\citenamefont{Read and Green}(2000)}]{Read-2000}
\bibinfo{author}{\bibfnamefont{N.}~\bibnamefont{Read}} \bibnamefont{and}
  \bibinfo{author}{\bibfnamefont{D.}~\bibnamefont{Green}},
  \bibinfo{journal}{Phys. Rev. B} \textbf{\bibinfo{volume}{61}},
  \bibinfo{pages}{10267} (\bibinfo{year}{2000}).

\bibitem[{\citenamefont{Fu and Kane}(2008)}]{Fu-2008}
\bibinfo{author}{\bibfnamefont{L.}~\bibnamefont{Fu}} \bibnamefont{and}
  \bibinfo{author}{\bibfnamefont{C.}~\bibnamefont{Kane}},
  \bibinfo{journal}{Phys. Rev. Lett.} \textbf{\bibinfo{volume}{100}},
  \bibinfo{pages}{096407} (\bibinfo{year}{2008}).

\bibitem[{\citenamefont{Lutchyn et~al.}(2010)\citenamefont{Lutchyn, Sau, and
  Das~Sarma}}]{Lutchyn-2010}
\bibinfo{author}{\bibfnamefont{R.~M.} \bibnamefont{Lutchyn}},
  \bibinfo{author}{\bibfnamefont{J.~D.} \bibnamefont{Sau}}, \bibnamefont{and}
  \bibinfo{author}{\bibfnamefont{S.}~\bibnamefont{Das~Sarma}},
  \bibinfo{journal}{Phys. Rev. Lett.} \textbf{\bibinfo{volume}{105}},
  \bibinfo{pages}{077001} (\bibinfo{year}{2010}).

\bibitem[{\citenamefont{Oreg et~al.}(2010)\citenamefont{Oreg, Refael, and von
  Oppen}}]{Oreg-2010}
\bibinfo{author}{\bibfnamefont{Y.}~\bibnamefont{Oreg}},
  \bibinfo{author}{\bibfnamefont{G.}~\bibnamefont{Refael}}, \bibnamefont{and}
  \bibinfo{author}{\bibfnamefont{F.}~\bibnamefont{von Oppen}},
  \bibinfo{journal}{Phys. Rev. Lett.} \textbf{\bibinfo{volume}{105}},
  \bibinfo{pages}{177002} (\bibinfo{year}{2010}).

\bibitem[{\citenamefont{Mourik et~al.}(2012)\citenamefont{Mourik, Zuo, Frolov,
  Plissard, Bakkers, and Kouwenhoven}}]{Mourik-2012}
\bibinfo{author}{\bibfnamefont{V.}~\bibnamefont{Mourik}},
  \bibinfo{author}{\bibfnamefont{K.}~\bibnamefont{Zuo}},
  \bibinfo{author}{\bibfnamefont{S.}~\bibnamefont{Frolov}},
  \bibinfo{author}{\bibfnamefont{S.}~\bibnamefont{Plissard}},
  \bibinfo{author}{\bibfnamefont{E.}~\bibnamefont{Bakkers}}, \bibnamefont{and}
  \bibinfo{author}{\bibfnamefont{L.}~\bibnamefont{Kouwenhoven}},
  \bibinfo{journal}{Science} \textbf{\bibinfo{volume}{336}},
  \bibinfo{pages}{1003} (\bibinfo{year}{2012}).

\bibitem[{\citenamefont{Deng et~al.}(2012)\citenamefont{Deng, Yu, Huang,
  Larsson, and Xu}}]{Deng-2012}
\bibinfo{author}{\bibfnamefont{M.}~\bibnamefont{Deng}},
  \bibinfo{author}{\bibfnamefont{C.}~\bibnamefont{Yu}},
  \bibinfo{author}{\bibfnamefont{G.}~\bibnamefont{Huang}},
  \bibinfo{author}{\bibfnamefont{M.}~\bibnamefont{Larsson}}, \bibnamefont{and}
  \bibinfo{author}{\bibfnamefont{H.}~\bibnamefont{Xu}}, \bibinfo{journal}{Nano
  Letters} \textbf{\bibinfo{volume}{12}}, \bibinfo{pages}{6414}
  (\bibinfo{year}{2012}).

\bibitem[{\citenamefont{Das et~al.}(2012)\citenamefont{Das, Ronen, Most, Oreg,
  Heiblum, and Shtrikman}}]{Das-2012}
\bibinfo{author}{\bibfnamefont{A.}~\bibnamefont{Das}},
  \bibinfo{author}{\bibfnamefont{Y.}~\bibnamefont{Ronen}},
  \bibinfo{author}{\bibfnamefont{Y.}~\bibnamefont{Most}},
  \bibinfo{author}{\bibfnamefont{Y.}~\bibnamefont{Oreg}},
  \bibinfo{author}{\bibfnamefont{M.}~\bibnamefont{Heiblum}}, \bibnamefont{and}
  \bibinfo{author}{\bibfnamefont{H.}~\bibnamefont{Shtrikman}},
  \bibinfo{journal}{Nat. Phys.} \textbf{\bibinfo{volume}{8}},
  \bibinfo{pages}{887} (\bibinfo{year}{2012}).

\bibitem[{\citenamefont{Churchill et~al.}(2013)\citenamefont{Churchill, Fatemi,
  Grove-Rasmussen, Deng, Caroff, Xu, and Marcus}}]{Churchill-2013}
\bibinfo{author}{\bibfnamefont{H.}~\bibnamefont{Churchill}},
  \bibinfo{author}{\bibfnamefont{V.}~\bibnamefont{Fatemi}},
  \bibinfo{author}{\bibfnamefont{K.}~\bibnamefont{Grove-Rasmussen}},
  \bibinfo{author}{\bibfnamefont{M.}~\bibnamefont{Deng}},
  \bibinfo{author}{\bibfnamefont{P.}~\bibnamefont{Caroff}},
  \bibinfo{author}{\bibfnamefont{H.}~\bibnamefont{Xu}}, \bibnamefont{and}
  \bibinfo{author}{\bibfnamefont{C.}~\bibnamefont{Marcus}},
  \bibinfo{journal}{Phys. Rev. B} \textbf{\bibinfo{volume}{87}},
  \bibinfo{pages}{241401(R)} (\bibinfo{year}{2013}).

\bibitem[{\citenamefont{Bolech and Demler}(2007)}]{Bolech-2007}
\bibinfo{author}{\bibfnamefont{C.}~\bibnamefont{Bolech}} \bibnamefont{and}
  \bibinfo{author}{\bibfnamefont{E.}~\bibnamefont{Demler}},
  \bibinfo{journal}{Phys. Rev. Lett.} \textbf{\bibinfo{volume}{98}},
  \bibinfo{pages}{237002} (\bibinfo{year}{2007}).

\bibitem[{\citenamefont{Law et~al.}(2009)\citenamefont{Law, Lee, and
  Ng}}]{Law-2009}
\bibinfo{author}{\bibfnamefont{K.}~\bibnamefont{Law}},
  \bibinfo{author}{\bibfnamefont{P.~A.} \bibnamefont{Lee}}, \bibnamefont{and}
  \bibinfo{author}{\bibfnamefont{T.}~\bibnamefont{Ng}}, \bibinfo{journal}{Phys.
  Rev. Lett.} \textbf{\bibinfo{volume}{103}}, \bibinfo{pages}{237001}
  (\bibinfo{year}{2009}).

\bibitem[{\citenamefont{Liu et~al.}(2012)\citenamefont{Liu, Potter, Law, and
  Lee}}]{Liu-2012}
\bibinfo{author}{\bibfnamefont{J.}~\bibnamefont{Liu}},
  \bibinfo{author}{\bibfnamefont{A.~C.} \bibnamefont{Potter}},
  \bibinfo{author}{\bibfnamefont{K.}~\bibnamefont{Law}}, \bibnamefont{and}
  \bibinfo{author}{\bibfnamefont{P.~A.} \bibnamefont{Lee}},
  \bibinfo{journal}{Phys. Rev. Lett.} \textbf{\bibinfo{volume}{109}},
  \bibinfo{pages}{267002} (\bibinfo{year}{2012}).

\bibitem[{\citenamefont{Nayak et~al.}(2008)\citenamefont{Nayak, Simon, Stern,
  Freedman, and Das~Sarma}}]{Nayak-2008}
\bibinfo{author}{\bibfnamefont{C.}~\bibnamefont{Nayak}},
  \bibinfo{author}{\bibfnamefont{S.~H.} \bibnamefont{Simon}},
  \bibinfo{author}{\bibfnamefont{A.}~\bibnamefont{Stern}},
  \bibinfo{author}{\bibfnamefont{M.}~\bibnamefont{Freedman}}, \bibnamefont{and}
  \bibinfo{author}{\bibfnamefont{S.}~\bibnamefont{Das~Sarma}},
  \bibinfo{journal}{Rev. Mod. Phys.} \textbf{\bibinfo{volume}{80}},
  \bibinfo{pages}{1083} (\bibinfo{year}{2008}).

\bibitem[{\citenamefont{Rainis and Loss}(2012)}]{Rainis-2012}
\bibinfo{author}{\bibfnamefont{D.}~\bibnamefont{Rainis}} \bibnamefont{and}
  \bibinfo{author}{\bibfnamefont{D.}~\bibnamefont{Loss}},
  \bibinfo{journal}{Phys. Rev. B} \textbf{\bibinfo{volume}{85}},
  \bibinfo{pages}{174533} (\bibinfo{year}{2012}).

\bibitem[{\citenamefont{Fu and Kane}(2009)}]{Fu-2009}
\bibinfo{author}{\bibfnamefont{L.}~\bibnamefont{Fu}} \bibnamefont{and}
  \bibinfo{author}{\bibfnamefont{C.}~\bibnamefont{Kane}},
  \bibinfo{journal}{Phys. Rev. B} \textbf{\bibinfo{volume}{79}},
  \bibinfo{pages}{161408(R)} (\bibinfo{year}{2009}).

\bibitem[{\citenamefont{Leijnse and Flensberg}(2011)}]{Leijnse-2011}
\bibinfo{author}{\bibfnamefont{M.}~\bibnamefont{Leijnse}} \bibnamefont{and}
  \bibinfo{author}{\bibfnamefont{K.}~\bibnamefont{Flensberg}},
  \bibinfo{journal}{Phys. Rev. b} \textbf{\bibinfo{volume}{84}},
  \bibinfo{pages}{140501(R)} (\bibinfo{year}{2011}).

\bibitem[{\citenamefont{Burnell et~al.}(2013)\citenamefont{Burnell, Shnirman,
  and Oreg}}]{Burnell-2013}
\bibinfo{author}{\bibfnamefont{F.}~\bibnamefont{Burnell}},
  \bibinfo{author}{\bibfnamefont{A.}~\bibnamefont{Shnirman}}, \bibnamefont{and}
  \bibinfo{author}{\bibfnamefont{Y.}~\bibnamefont{Oreg}},
  \bibinfo{journal}{arXiv:1206.6687}  (\bibinfo{year}{2013}).

\bibitem[{\citenamefont{Nilsson et~al.}(2008)\citenamefont{Nilsson, Akhmerov,
  and Beenakker}}]{Nilsson-2008}
\bibinfo{author}{\bibfnamefont{J.}~\bibnamefont{Nilsson}},
  \bibinfo{author}{\bibfnamefont{A.}~\bibnamefont{Akhmerov}}, \bibnamefont{and}
  \bibinfo{author}{\bibfnamefont{C.}~\bibnamefont{Beenakker}},
  \bibinfo{journal}{Phys. Rev. Lett.} \textbf{\bibinfo{volume}{101}}
  (\bibinfo{year}{2008}).

\bibitem[{\citenamefont{Liu et~al.}(2013)\citenamefont{Liu, Zhang, and
  Law}}]{Liu-2013}
\bibinfo{author}{\bibfnamefont{J.}~\bibnamefont{Liu}},
  \bibinfo{author}{\bibfnamefont{F.}~\bibnamefont{Zhang}}, \bibnamefont{and}
  \bibinfo{author}{\bibfnamefont{K.}~\bibnamefont{Law}},
  \bibinfo{journal}{Phys. Rev. B} \textbf{\bibinfo{volume}{88}},
  \bibinfo{pages}{064509} (\bibinfo{year}{2013}).

\bibitem[{\citenamefont{Anantram and Datta}(1996)}]{Anantram-1996}
\bibinfo{author}{\bibfnamefont{M.}~\bibnamefont{Anantram}} \bibnamefont{and}
  \bibinfo{author}{\bibfnamefont{S.}~\bibnamefont{Datta}},
  \bibinfo{journal}{Phys. Rev. B} \textbf{\bibinfo{volume}{53}},
  \bibinfo{pages}{16390} (\bibinfo{year}{1996}).

\end{thebibliography}
\section*{Appendix A: Finite Temperature Calculations }
To calculate the differential conductivity and noise correlations above zero temperature we used the relations given in Ref. \cite{Anantram-1996} given in our notation by
$$I_\alpha=\frac{e}{h}\sum_{\alpha,i,j}sgn(\alpha)\int dE[\delta_{\alpha\beta}\delta_{ij}-|S_{\alpha,\beta,ij}|^2]f_{\alpha,j}(E)$$
\begin{align*}
C_{\alpha\beta}=& \frac{2e^2}{h}\sum_{\gamma,\delta,i,j,k,l}\mathrm{sgn}(i)\mathrm{sgn}(j)\\
&\int dE A_{\gamma k;\delta l}(\alpha i,E)A_{\delta l;\gamma k}(\beta j,E)f_{\gamma,k}(E)[1-f_{\delta, l}(E)]
\end{align*}
where $\alpha,\beta,\gamma,\delta$ label the lead or reservoir and $i,j,k,l\in\{e,h\}$ label hole and electron channels. 
The functions $f$, sgn, and $A$ are given by $$f_{\alpha,j}(E)=\left[1+\mathrm{exp}\left(\frac{E-\mu_\alpha\mathrm{sgn}(j)}{k_BT}\right)\right]^{-1}$$ $\mathrm{sgn}(e)=1$ and $\mathrm{sgn}(h)=-1$
$$A_{\gamma k;\delta l}(\alpha i,E)=\delta_{\alpha\gamma}\delta_{\alpha\gamma}\delta_{ik}\delta_{il}-S_{\alpha,\gamma,ik}^*S_{\alpha,\delta,il}$$ $\mu_\alpha$ gives the chemical potential of the lead or reservoir compared to that of the superconductor. The only dependence on the voltage is given in the Fermi functions $f$ so the differential factors are given by 
$$G_\alpha=\frac{e^2}{h}\sum_{\alpha,i,j}sgn(\alpha)\int dE[\delta_{\alpha\beta}\delta_{ij}-|S_{\alpha,\beta,ij}|^2]\frac{df_{\alpha,j}}{dV}(E)$$ and 
\begin{align*}
P_{\alpha\beta}=& \frac{2e^2}{h}\sum_{\gamma,\delta,i,j,k,l}\mathrm{sgn}(i)\mathrm{sgn}(j)\\
&\times\int dE A_{\gamma k;\delta l}(\alpha i,E)A_{\delta l;\gamma k}(\beta j,E)\\
&\times\left(\frac{df_{\alpha,k}}{dV}(E)[1-f_{\delta,l}(E)]+f_{\alpha,k}(E)[1-\frac{df_{\delta,l}}{dV}(E)]\right)
\end{align*}

The temperature samples the differential conductance from an area of width $kT$. Away from resonance, where the differential conductance and noise change on the order of $\Gamma$, this sampling has little effect as the differential conductance and noise vary slowly in energy. 

\section*{Appendix B: Calculation of the Poisoning Rate}
The measurements given above do not give enough information to determine the poisoning rate, but if supplemented by the ratio of the height of the resonance of each lead, do allow its determination. Expanding the differential cross-correlation at low bias we get
\begin{equation}
P_{LR}=\frac{2e^2}{h}\frac{\Gamma_t^L\Gamma_t^R}{E_M^2}
\end{equation}
to lowest order in $\Gamma^2/E_M^2$ and $E^2/E_M^2$. The value of the Majorana splitting, $E_M$, is easily determined by the location of the resonance peak. From this we can calculate the product
\begin{equation}
\Gamma_t^L\Gamma_t^R=\frac{E_M^2h}{2e^2}P_{LR}
\end{equation} From equation \ref{eq:LowBiasConductance} above we can write the product 
\begin{equation}
\begin{aligned}
\Gamma_t^L\Gamma_p^R =& \frac{E_M^2h}{2e^2}G_L(eV=0)-\Gamma_t^L\Gamma_t^R \\
&= \frac{E_M^2h}{2e^2}\big(G_L(eV=0)-P_{LR}(eV=0)\big)
\end{aligned}
\label{eq:Product}
\end{equation}
and similarly for $\Gamma_t^R\Gamma_p^L$.

To measure the poisoning rate, $\Gamma_p^L+\Gamma_p^R$, we also need to know the ratio between tunneling on the two sides, $r=\Gamma_t^R/\Gamma_t^L$. $r$ can be measured by comparing the height of the differential conductance peak of each lead. The height of the resonance for $G_\alpha$ at zero temperature is given by $$\frac{2e^2}{h}\frac{\Gamma_t^\alpha}{\Gamma_t^L+\Gamma_t^R+\Gamma_p^L+\Gamma_p^R}$$ so taking the ratio of the conductance for each lead gives 
\begin{equation}
r=\frac{G_R(eV=E_M)}{G_L(eV=E_M)}
\end{equation} Even at temperatures $k_BT>\Gamma_\alpha$ the result holds because the thermal sampling of the differential conductance is dominated by the region near resonance where the result holds. 

Now in terms of these measurements we can write the sum of the poisoning rate as:
\begin{widetext}
\begin{equation}
\begin{aligned}
\Gamma_p^L+\Gamma_p^R=& \frac{1}{\sqrt{\Gamma_t^L\Gamma_t^R}} \left( \Gamma_p^L\Gamma_t^R\sqrt{\frac{1}{r}}+\Gamma_p^R\Gamma_t^L\sqrt{r} \right) \\
=&\sqrt{\frac{h}{2e^2}}\frac{1}{\sqrt{P_{LR}}}\left((G_R-P_{LR})\sqrt{\frac{1}{r}}+(G_L-P_{LR})\sqrt{r} \right)E_M
\end{aligned}
\label{eq:Rate}
\end{equation}
\end{widetext}
Multiplying Eq. \ref{eq:Rate} by $\Gamma_t^L$ and combining with Eq. \ref{eq:Product} allows us to determine $\Gamma_t^L\Gamma_p^L$. Combining with Eq. \ref{eq:Product} gives us the ratio $\Gamma_p^R/\Gamma_p^L$. Together with Eq. \ref{eq:Rate}, $\Gamma_p^R$ and $\Gamma_p^L$ are determined separately.
\end{document}